\documentclass{iau} 
\usepackage{graphicx}
\usepackage{natbib}
\usepackage{mathtools}

\pubyear{2022}
\volume{370}
\jname{Winds of Stars and Exoplanets}
\editors{A.~A. Vidotto, L.~Fossati \& J.~Vink, eds.}

\title[The Driving of Hot Star Winds] 
{The Driving of Hot Star Winds}

\author[A.A.C. Sander]   
{Andreas A.C. Sander$^1$}

\affiliation{$^1$Zentrum für Astronomie der Universität Heidelberg, Astronomisches Rechen-Institut, Mönchhofstr. 12-14, 69120 Heidelberg, Germany \\ email: {\tt andreas.sander@uni-heidelberg.de}}

\begin{document}

\maketitle

\begin{abstract}
In the regime of hot stars, winds were not seen as a common thing until the era of UV astronomy. Since we have access to the UV wavelength range, it has become clear that winds are not an exotic phenomenon limited to some special objects, but actually ubiquitous among hot and massive stars. The opacities due to spectral lines are the decisive ingredient that allows hot, massive stars to launch powerful winds. While the fundamental principles of these so-called line-driven winds have been realized decades ago, their proper quantitative prediction is still a major challenge today. Established theoretical and empirical descriptions have allowed us to make major progress on all astrophysical scales. However, we are now reaching their limitations as we still lack various fundamental insights on the nature of hot star winds, thereby hampering us from drawing deeper conclusions, not least when dealing with stellar or sub-stellar companions. This has spawned a new generation of researchers searching for answers with a yet unprecedented level of detail in observational and new theoretical approaches.

In these proceedings, the fundamental principles of driving hot star winds will be briefly reviewed. Starting from the classical CAK theory and its extensions, over Monte Carlo and recent comoving-frame-based simulations, the different methods to describe and model the acceleration of hot star winds will be introduced. The review continues with briefly discussing instabilities as well as qualitative and quantitative insights for OB- and Wolf-Rayet-star winds. Moreover, the challenges of companions and their impact on radiation-driven winds are outlined.

\end{abstract}

\firstsection 
              
\section{Introduction}

While hot stars ($T_\mathrm{eff} > 10\,$kK) appear blue to the human eye, their actual flux maximum is located at ultraviolet (UV) wavelengths. The onset of UV astronomy at the end of the 1960s \citep{Morton1967,LucySolomon1967} revealed that this highly energetic flux does not simply escape from the stars, but can give rise to strong stellar winds due to the absorption of the star's radiation in spectral lines. Albeit the photon is eventually re-emitted, this re-emission happens in an arbitrary direction, while most of the absorption occurs in radial direction. This results in a radial net transfer of momentum from the photons to the matter in the outermost layers of the star, giving rise to a stellar wind.  

The idea of radiation pressure being powerful enough to remove material from a stellar surface had already been proposed a generation earlier for known (optical) emission-line stars \citep{Milne1926,Beals1929}, but the ubiquitous occurrence of radiation-driven winds in hot stars was only discovered with the accessibility of the UV spectral range, revealing P\,Cygni profiles in many hot star spectra. Albeit hot star winds are subject to inherent instabilities \citep[e.g.,][see also Sect.\,\ref{sec:instabilities}]{LucySolomon1970}, the overall shape of the wind-affected line profiles is largely constant over time, thereby justifying also a stationary, i.e. time-independent, description. In this limit, the radiative acceleration for a spherically symmetric star can be written as 
					 \begin{equation}   
					     \label{eq:arad}
			  	  	 a_\mathrm{rad}(r) = \frac{1}{c} \int\limits_0^{\infty} \varkappa_\nu(r) F_\nu(r) \mathrm{d}\nu = {\varkappa_F(r)} \frac{L}{4 \pi c r^2}\mathrm{.}
				   \end{equation}

It is usually convenient to express $a_\mathrm{rad}$ not as an absolute value, but normalize it to the gravitational acceleration that tries to pull material back to the star. One can thus define the quantity
  \begin{equation}    
	  \label{eq:gamrad}
	 	\Gamma_\mathrm{rad}(r) := \frac{a_\mathrm{rad}(r)}{g(r)} = {\varkappa_F(r)} \frac{L}{4 \pi c G {M}}\mathrm{.}
	\end{equation}
From Eq.\,(\ref{eq:gamrad}), we immediately obtain that the strength of a radiatively driven wind depends mainly on three quantities: the stellar luminosity $L$, its mass $M$, and the \emph{flux-weighted mean opacity} $\varkappa_F(r)$. The latter opacity dependence is what makes the description of radiatively-driven winds complicated as $\varkappa_F(r)$ has a multitude of inherent dependencies on the physical conditions in the outermost layers of a star.

\begin{figure}[htb]
\begin{center}
   \includegraphics[angle=0,width=0.49\textwidth]{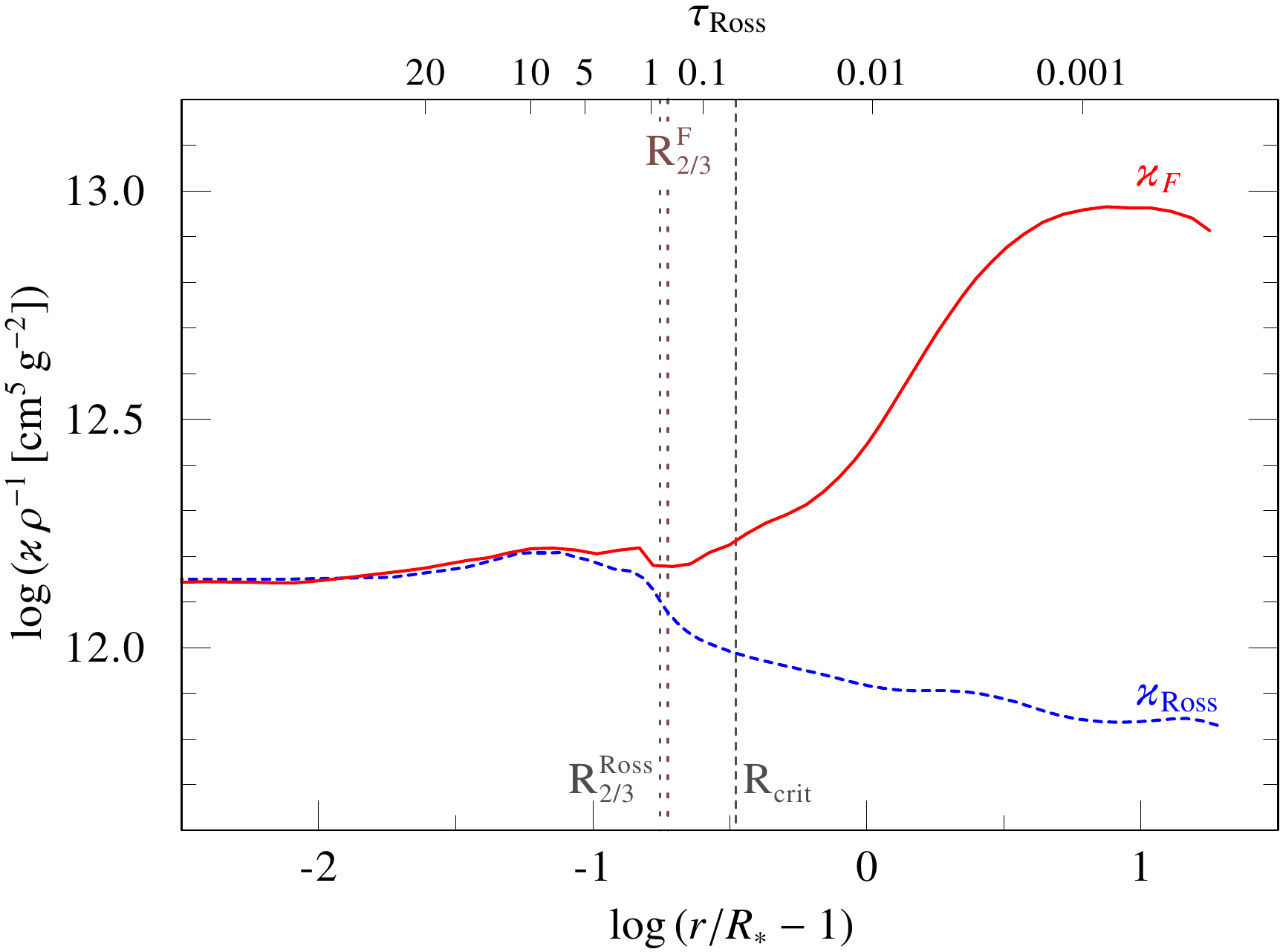} \hfill
   \includegraphics[angle=0,width=0.49\textwidth]{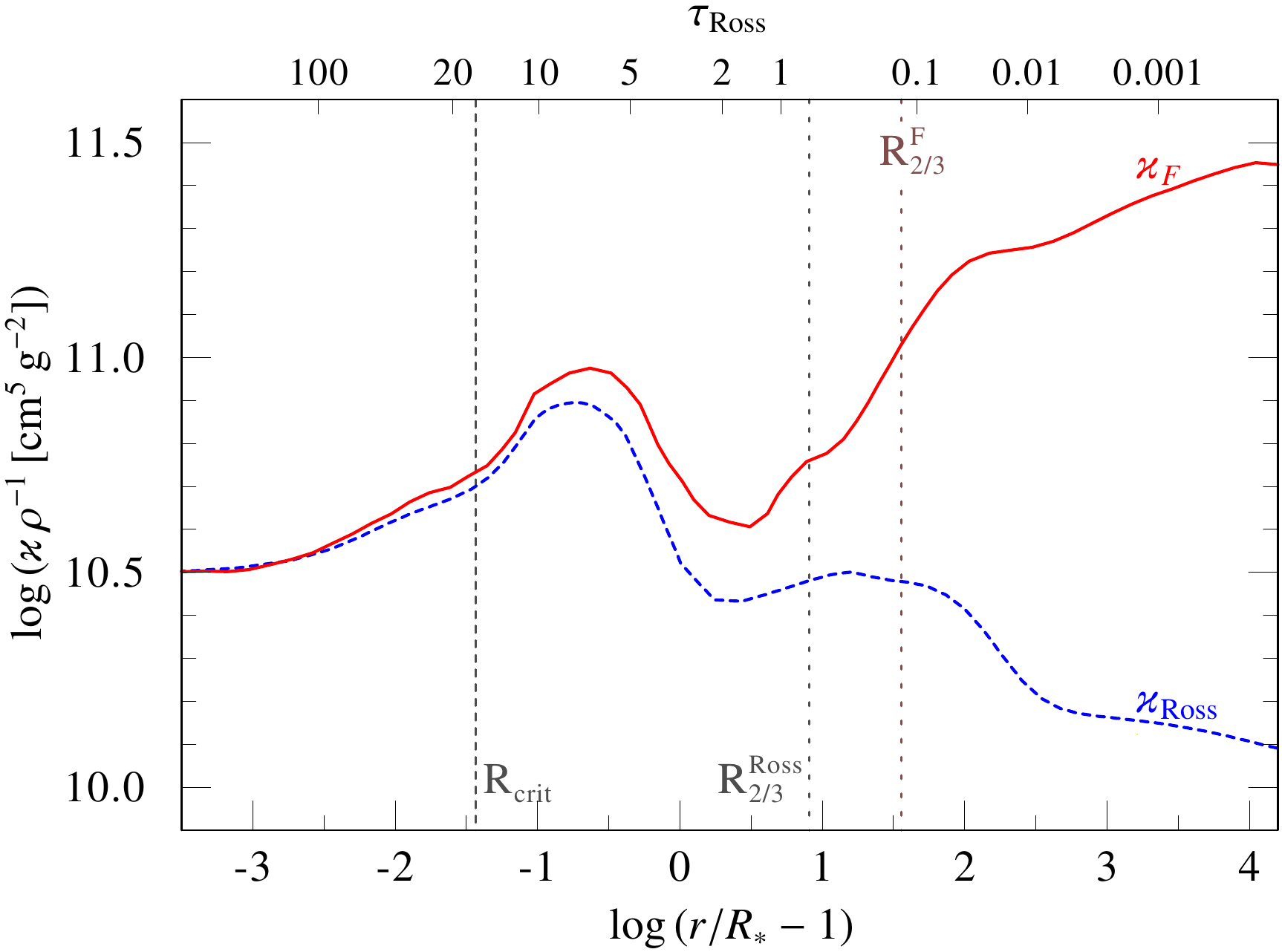}
     \caption{Comparison of flux-weighted (red, solid) and Rosseland mean opacity (blue, dashed) for dynamically-consistent atmosphere models of a B supergiant (left panel) and a hydrogen-free WN star (right panel).}
   \label{fig:checkopa}
\end{center}
\end{figure}

The flux-weighted mean opacity $\varkappa_F$ is crucial to understand how stellar winds can escape a star, even if they are considerably below the classical  Eddington Limit of $\Gamma_\mathrm{e} = 1$, defining the traditional structural stability limit for a star. The definition of $\Gamma_\mathrm{e}$ is similar to Eq.\,(\ref{eq:gamrad}), but using only the Thomson opacity $\varkappa_\mathrm{Thomson} \rho(r) = \sigma_\mathrm{e} n_\mathrm{e}(r)$ describing the scattering of free electrons instead of the full $\varkappa_F$. Although $\varkappa_\mathrm{Thomson}$ usually makes a large contribution to the radiative acceleration $a_\mathrm{rad}$ of a stellar wind, it would (in most cases) not be sufficient to overcome gravity on its own. Instead, the aforementioned absorption in spectral lines plays a key role as a second important component. Bound-bound and bound-free opacities can contribute as well, resulting in a total opacity of 
\begin{equation}
  \label{eq:opatot}
  \varkappa = \varkappa_\mathrm{bound-free} + \varkappa_\mathrm{free-free} + \varkappa_\mathrm{Thomson} + \varkappa_\mathrm{lines}
\end{equation}
for hot star winds where components such as molecular transitions or dust are absent due to the high $T_\mathrm{eff}$-regime. 

Given that the calculation of the components beyond $\varkappa_\mathrm{Thomson}$ in Eq.\,(\ref{eq:opatot}) requires a significant numerical effort, the use of tabulated opacities such as OPAL \citep{IglesiasRogers1996} is a common ingredient in astrophysical simulations, e.g., in stellar evolution modelling or time-dependent (magneto-)hydrodynamical simulations. Such tables provide the Rosseland opacity
\begin{equation}
  \label{eq:opaross}
  \varkappa_\mathrm{Ross}^{-1} = \frac{ \int_{0}^{\infty} \varkappa_{\nu}^{-1} \frac{\partial B_\nu}{\partial T} \mathrm{d}\nu }{ \int_{0}^{\infty} \frac{\partial B_\nu}{\partial T} \mathrm{d}\nu }
\end{equation}
as a function of temperature and density. As evident from comparing Eq.\,(\ref{eq:opaross}) with the implicit definition of $\varkappa_F$ in Eq.\,(\ref{eq:arad}), the definition of the Rosseland opacity $\varkappa_\mathrm{Ross}$ and the flux-weighted mean opacity $\varkappa_F$ is not identical. As long as we are in an optically thick regime that fulfils the conditions of local thermodynamic equilibrium (LTE), the values of $\varkappa_F$ can be approximated with $\varkappa_\mathrm{Ross}$. The winds of hot stars, however, are not in LTE. In Fig.\,\ref{fig:checkopa}, the difference between the two quantities is illustrated for two different types of hot stars. In the inner, hydrostatic regime where the star is optically thick, the two quantities align. Further out in the stellar wind, $\varkappa_F$ considerably exceeds the values of $\varkappa_\mathrm{Ross}$, revealing a completely different trend due to the Doppler-shift of the spectral lines in the wind, giving rise to additional absorption (and thus opacity) at blueshifted wavelengths. Consequently, simulations relying (only) on Rosseland opacity tables cannot produce a realistic stellar wind situation.

\section{Methods to calculate the radiative acceleration}
  \label{sec:concepts}

With the discovery of the UV P\,Cygni profiles, the principles of radiative driving were quickly inferred, providing first formulations of hot star mass-loss and  theoretical considerations for radiation-driven winds \citep[e.g.][]{LucySolomon1970,Castor1974}. These efforts eventually culminated in the work by Castor, Abbott, \& Klein (1975) \nocite{Castor+1975} giving an analytic approximation for $a_\mathrm{rad}$ and the resulting mass-loss rate $\dot{M}$. In what is nowadays called the \emph{CAK theory} -- abbreviated after their initials -- one assumes that the contributions from bound-free and free-free opacity are negligible. Hence, Eq.\,(\ref{eq:opatot}) reduces to:
\begin{equation}
  \label{eq:opacak}
  \varkappa \approx \varkappa_\mathrm{Thom} + \varkappa_\mathrm{lines} = (1+\mathcal{M}) \varkappa_\mathrm{Thom}\mathrm{.}
\end{equation}
In the second formulation of Eq.\,(\ref{eq:opacak}), the line opacities are expressed in the form of the Thomson opacity, multiplied with a so-called \emph{force multiplier} $\mathcal{M}$. The computation of $\mathcal{M}$ can be further simplified by considering only photons emerging radially from the stellar disk (``radial streaming approximation'') and using the Sobolev approximation \citep{Sobolev1960}. Neglecting line overlap and splitting strictly into optically thin and thick cases, $\mathcal{M}$ can be calculated as
\begin{equation}
  \mathcal{M}(t) = \frac{1}{c} \sum\limits_\mathrm{i=1}^{N_\mathrm{lines}} \left(\Delta\nu_\mathrm{D} F_\nu\right)_i 
	 \begin{cases} 
	    \varkappa_\mathrm{line} / \varkappa_\mathrm{Thom} & \text{for optically thin lines} \\
	    t^{-1} & \text{for optically thick lines} \\
	 \end{cases}
\end{equation}
with $t$ denoting the so-called Sobolev optical depth $t = \varkappa_\mathrm{e} \rho v_\mathrm{th} \left| \frac{\mathrm{d}v}{\mathrm{d}r} \right|^{-1}$, which is independent of a particular line opacity. The resulting curve for $\mathcal{M}(t)$ is then approximated as
\begin{equation}
  \label{eq:forcemultcak}
  \mathcal{M} = k t^{-\alpha} \left( 10^{-11} \mathrm{cm}^{3} \frac{n_\mathrm{e}(r)}{W(r)} \right)^{\delta}\mathrm{.}
\end{equation}
The expression (\ref{eq:forcemultcak}) for $\mathcal{M}$ employs already the extended notation with three coefficients $k$, $\alpha$, $\delta$, the latter including the geometrical dilution factor $W(r)$. While $\alpha$ and $k$ go back to the original CAK work, the $\delta$-parameter was introduced by \citet{Abbott1982} to account for ionization changes in the wind. The CAK descriptions were then further extended in the following two decades \citep[e.g.][]{FriendAbbott1986,Pauldrach+1986,Kudritzki+1989,Puls+2000}, most notably to remove the radial streaming approximation and to interpret the connection to the underlying line-strength distribution. It is the simplicity of requiring only a power-law to describe $a_\mathrm{rad}$ -- and a set of coefficients -- that explains the success of the (modified) CAK theory until today. The solution of the equation of motion using the CAK description for $a_\mathrm{rad}$ also motivates the $\beta$-velocity law 
\begin{equation}
  v(r) = v_\infty \left( 1 - \frac{R_\ast}{r} \right)^\beta
\end{equation}
to describe the velocity stratification of a hot star wind. In the original CAK calculation with negligible gas pressure (``zero sound-speed approximation''), one obtains $\beta = 0.5$. With further considerations and later extensions, values of $\beta \approx 0.8$ are nowadays seen as more representative. In the quantitative spectroscopy of hot stars with model atmospheres, $\beta$ is often a free parameter that can be indirectly constrained from reproducing UV line profiles together with other spectral features. 

An alternative approaches to obtain the $a_\mathrm{rad}$ is to compute the radiative transfer using a Monte Carlo (MC) approach. For hot star winds, this method was first implemented by \citet{AbbottLucy1985}. Technically, the interaction of so-called ``photon packages'' are tracked throughout the wind, which is described as a series of layers (1D shells). In each shell, photons can potentially transfer momentum and energy to the wind material. In contrast to CAK, this method can account for multiple scatterings of each photon, which becomes important in more dense winds that are characterized by higher mass-loss rates. The dense winds of Wolf-Rayet (WR) stars, which cannot be explained in the framework of CAK, have thus been one of the focus applications of MC radiative transfer codes \citep[e.g.][]{Schmutz1993,deKoter+1997,VinkdeKoter2005}. The known loss of energy and thus luminosity between the onset of the wind and the outer boundary provides a way to compute the mass-loss rate $\dot{M}$ via the relation
\begin{equation}
  \label{eq:mdotmc}
  \frac{1}{2} \dot{M} \left(v_\infty - v_\mathrm{esc}\right) = L(R_\ast) - L(r \rightarrow \infty)
\end{equation}
if the terminal wind velocity $v_\infty$ is known, e.g.\ by assuming a $\beta$-law. The widely used mass-loss recipe for OB stars from \citet{Vink+2000,Vink+2001} makes use of this, assuming a scaling of $v_\infty$ with $v_\mathrm{esc}$ and a $\beta$-law. Later, \citet{MuellerVink2008} extended the method to also predict $v_\infty$. The same technique is used in more recent MC calculations such as in \citet{VinkSander2021}. Combined with some approximations for the atmosphere, the 1D MC calculations are relatively fast on modern computers and have thus been used to probe a wider parameter space with various grids of models. 
 A major outcome of these calculations was the association of an observed change in the terminal velocities around $T_\text{eff} \approx 21\,$kK \citep{Lamers+1995} with a change in mass-loss rates \citep{Vink+1999}. This increase in $\dot{M}$ towards cooler temperatures is termed the \emph{bi-stability jump}, using a terminology introduced originally in modelling efforts for the B hypergiant P\,Cyg \citep{PauldrachPuls1990}. MC models further helped to shape our understanding about the different role of iron (and iron-group elements) compared to CNO and intermediate-mass elements in O-star winds with the first group determining the conditions of the inner wind (and thus $\dot{M}$) and the second group providing the acceleration in the outer wind, thereby setting the terminal velocity $v_\infty$ \citep[e.g.][]{deKoter+1997,Vink+1999}.

Given its straight-forward principles, the MC concept can also be used for multi-dimensional simulations. While this is computationally more costly and thus has not been used for large model grids so far, \citet{Surlan+2012,Surlan+2013} have used the concept to model the effect of optically thick clumps on the spectral imprint on UV resonance lines in a hot star wind.

Despite the success of the MC approach, it has also limitations. From Eq.\,(\ref{eq:mdotmc}), there is no guarantee that the hydrodynamic equation of motion
\begin{equation}
  \label{eq:eom}
  a_\mathrm{rad} - \frac{1}{\rho} \frac{\mathrm{d}P_\mathrm{gas}}{\mathrm{d}r} = v \frac{\mathrm{d}v}{\mathrm{d}r} + g(r)
\end{equation} 
describing the balance of radiation and gas pressure versus gravity and inertia is actually fulfilled locally, i.e.\ at every point in the stellar wind. Instead, only the global energy budget, which can be obtained from integrating Eq.\,(\ref{eq:eom}), is consistent. The local consistency can be improved by reformulating the equation of motion in a way that the velocity field can be described by the LambertW function \citep[see, e.g.,][]{MuellerVink2008,Gormaz-Matamala+2021}, but requires the assumption of an isothermal wind. In particular more dense winds are not well approximated by a constant temperature, thereby limiting the validity of the MC results, in particular for the crucial regime of the wind launching region, which sets the mass-loss rate $\dot{M}$.

To obtain a locally consistent description of the radiative acceleration, a third approach has emerged that has traditionally been used in quantitative spectroscopy. The solution of the radiative transfer in the co-moving frame (CMF) goes back to a series of papers starting with \citet{Mihalas+1975}, describing what is essentially a numerical solution of the integral in Eq.\,(\ref{eq:arad}). While the opacities and emissivities do not remain isotropic in the observer's frame, they do so in the CMF, providing a significant advantage in programming and computing. This method has been widely applied in expanding stellar atmosphere codes for quantitative spectroscopy such as CMFGEN \citep[e.g.][]{HillierMiller1998}, PoWR \citep[e.g.][]{Hamann+1991}, FASTWIND \citep[e.g.][]{Puls+2020}, or PHOENIX \citep[e.g.][]{BaronHauschildt1998}. With their focus on spectral analysis, the elemental coverage was traditionally limited to elements that either were visible in the observable part of the spectrum or necessary for obtaining a correct temperature and ionization equilibrium. Nonetheless, their local solution of $a_\mathrm{rad}(r)$ make CMF models a suitable input for a local solution of the hydrodynamic equation of motion (\ref{eq:eom}). 

While such hydrodynamically-consistent CMF calculations were envisioned already in the 1980s \citep[see][]{Pauldrach+1986}, their success only became possible with the capabilities to include many different elements and solve initial stability problems. The first successful model was presented by \citet{GraefenerHamann2005}, reproducing the spectrum of a prototypical early-type WC star. Later, a series of hydrogen-rich, late-type WN stars \citep{GraefenerHamann2008} yielded a first $\dot{M}$ recipe based on dynamically-consistent CMF atmosphere models. The method has since also been applied to O- and B-type stars with different codes varying in the implementation details \citep[e.g.][]{KrtickaKubat2010,Sander+2017,Sundqvist+2019}. For OB-type wind predictions, the CMF-based calculations \citep[e.g.][]{KrtickaKubat2017,KrtickaKubat2018,Bjoerklund+2021} yield a considerable reduction in mass-loss rates compared to the \citet{Vink+2001} description. First comparisons with observations \citep[e.g.][]{Hawcroft+2021,Ramachandran+2019} indicate that the inferred rates from OB-type models could be too low, at least in some parameter regimes. Another issue, e.g.\ seen in OB-type models by \citet{Bjoerklund+2021}, are the often too high terminal velocities obtained in the CMF models. Identifying the underlying reasons is a matter of ongoing research and crucial to obtain better wind predictions. Recent models for Wolf-Rayet stars \citep{Sander+2020,SanderVink2020} do not show the same pattern. One possible ingredient could be the uncertain turbulent velocities at the base of the wind, which can have a strong influence on the derived $\dot{M}$ \citep[e.g.][]{Lucy2010,KrtickaKubat2017,Bjoerklund+2021} and likely differ between OB and WR-type stars. Moreover, there might be unaccounted processes in the 1D models, especially in the case of optically thin winds, resulting from multi-D effects. While a full CMF-based 3D wind treatment is computationally not feasible, ongoing efforts exist towards more realistic 3D simulations \citep[e.g.][]{Moens+2022,Poniatowski+2022} which could provide the necessary constraints and descriptions for 1D models with a more detailed radiative transfer.

The three methods to calculate the radiative acceleration (mCAK, MC, CMF) provide complementary toolsets, reaching from fast, but more approximate, to very detailed, but numerically expensive treatments. Their insights have enhanced our view on radiative driving, but have also prompted new questions. The recent progress in the CMF-based models is promising, but eventually the need for a CAK-like, parametrized description will be necessary for a realistic, but also efficient treatment of hot star winds in time-dependent and large-scale simulations.

\section{Instabilities and wind clumping}
  \label{sec:instabilities}
	
The acceleration mechanism of line-driven winds is subject to inherent instabilities. This was expected already by \citet{LucySolomon1970} when introducing their steady-state formalism to explain the observed OB-star winds. It took about a decade until multiple approaches aimed to describe the time-dependent behaviour and the resulting instabilities \citep[e.g.][]{Holzer1977,MacGregor+1979,Carlberg1980}. These were later combined in the line-force perturbation analysis work of \citet{OwockiRybicki1984}. The basic idea behind the instability of line-driven acceleration stems from the picture that a velocity-dependent line-shift enables new absorption capabilities. Due to the wind, a particular line transition can absorb photons of different wavelengths with bluer wavelengths being reached for higher wind velocities. Consequently, a positive perturbation in the velocity field exposes material to ``fresh'' flux. This in turn leads to additional absorption and thus further wind acceleration, thereby again enhancing the wind velocity. 

Various simulation efforts have been performed over the decades \citep[e.g.][]{Feldmeier+1997,OwockiPuls1999,SundqvistOwocki2013}, including extensions to multiple dimensions \citep[][]{DessartOwocki2005,Driessen+2022} or the consideration of magnetic fields \citep[][]{udDoulaOwocki2002,Driessen+2021}. To avoid extensive numerical costs or obtain (semi-)analytic descriptions, the use of the CAK theory and its later reformulations and extensions \citep[e.g.,][]{Gayley1995,OwockiPuls1999} is common. Still, it is important to underline that while the absolute magnitude of the instabilities in line-driven winds is sensitive to its physical approximations and numerical treatments, their existence as such is not a product of any simplifications. The so-called \emph{line-driven instability} or \emph{line-deshadowing instability}, abbreviated LDI, is expected from fundamental considerations.

The LDI provides a mechanism for breaking up homogeneous wind structures, causing porosity (i.e.\ ``clumping'') in physical and velocity space. Wind-intrinsic X-rays in hot star winds are commonly attributed to shocks resulting from colliding clumps \citep[][]{Feldmeier+1997}. Yet, it is unclear whether the LDI alone can fully explain the observed clumping and X-rays. There are observational \citep[e.g.,][]{LepineMoffat2008} and structural indications \citep[e.g.,][]{Cantiello+2009} that inhomogeneities exist already beneath the regime that is susceptible to the LDI. Consequently, the LDI might only enhance effects that occur already in the near-sonic layers of hot stars.

\section{The complex contribution of the different elements and ions}
  \label{sec:contributions}

Beside the acceleration due to free electrons, \citet{Castor+1975} included only C\,\textsc{iii} in their original paper. Due to their prominent P\,Cygni profiles in the UV, the CNO elements were initially seen as the main contributors for driving the winds of hot stars. Later extensions to more elements \citep[e.g.,][]{Abbott1982} then revealed that this picture was too simplified and also elements like iron (Fe) were playing an important role with different elements and ions dominating the wind driving in different temperature regimes. Moreover, the CAK-inherent approximation that bound-free and free-free opacities do not significantly contribute to $a_\mathrm{rad}$ turned out to be invalid for more dense winds.

\begin{figure}[htb]
\begin{center}
   \includegraphics[angle=0,width=\textwidth]{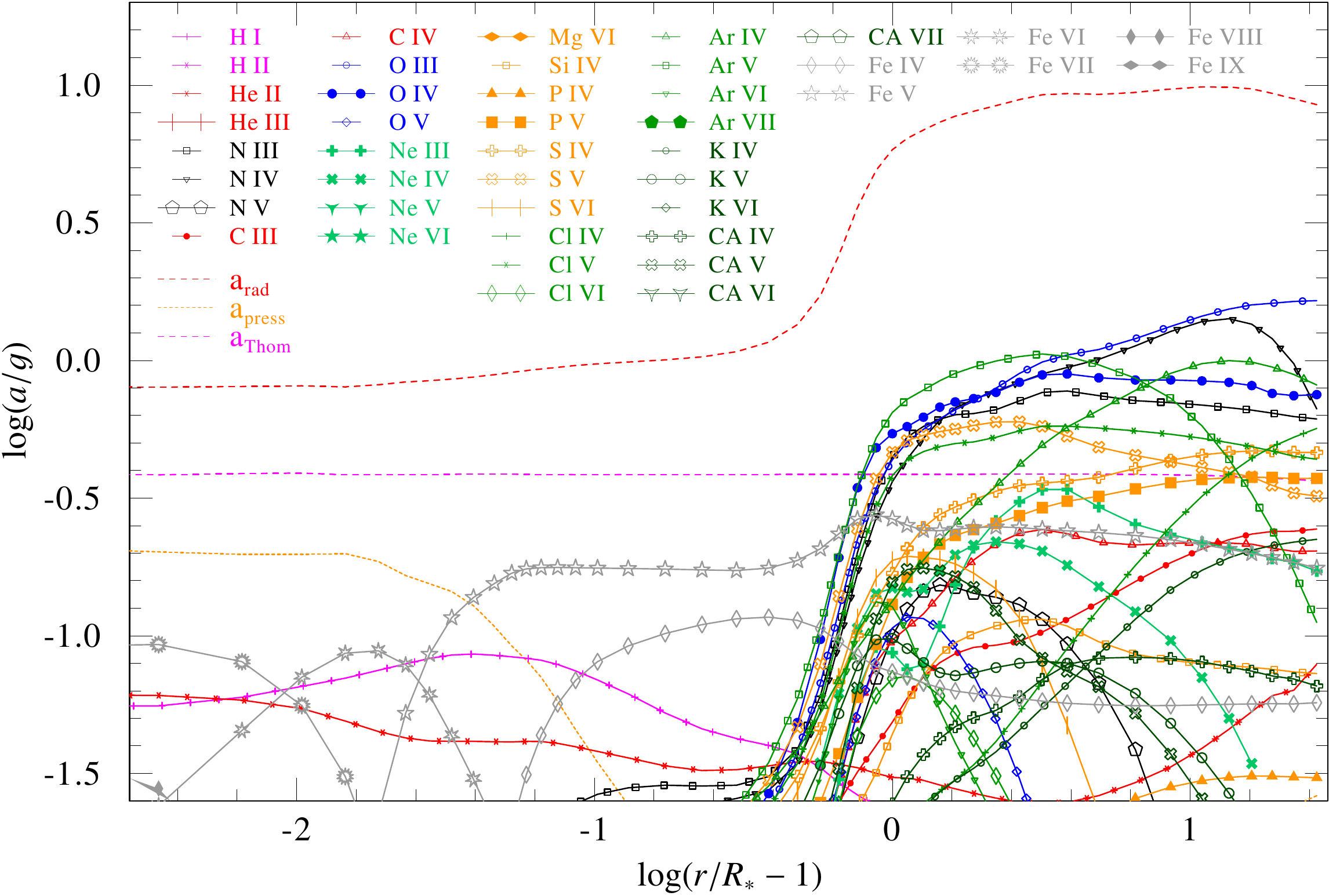}
     \caption{Contributions to the radiative acceleration in an O supergiant with $T_\mathrm{eff} = 42\,$kK.
		          The underlying model to produce this plot is taken from \citet{Sander+2017}.}
   \label{fig:leadions-osg}
\end{center}
\end{figure}

\begin{figure}[htb]
\begin{center}
   \includegraphics[angle=0,width=\textwidth]{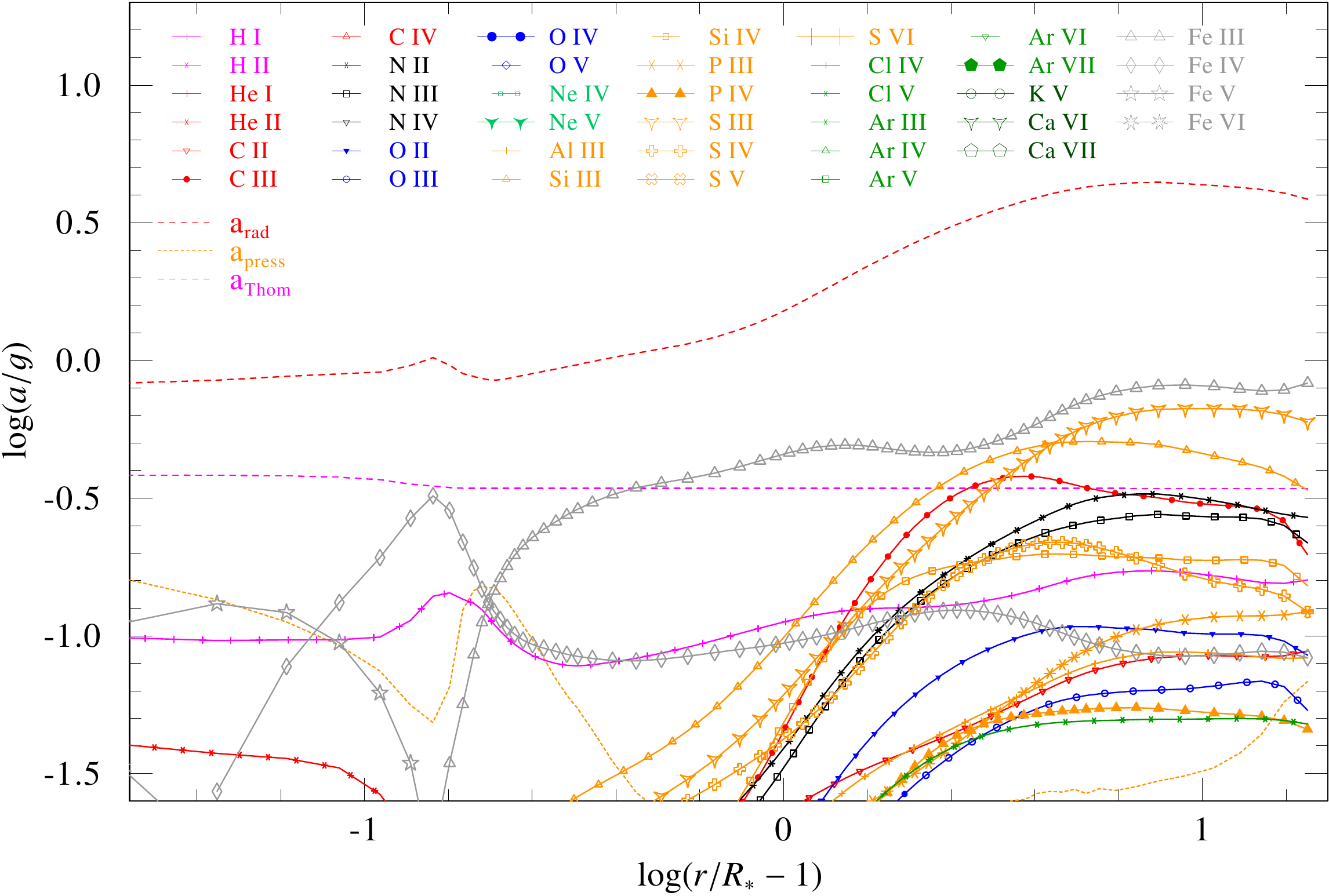}
     \caption{Contributions to the radiative acceleration in a B supergiant with $T_\mathrm{eff} = 25\,$kK.
		          The underlying model to produce this plot is taken from \citet{Sander+2018}.}
   \label{fig:leadions-bsg}
\end{center}
\end{figure}

\begin{figure}[htb]
\begin{center}
   \includegraphics[angle=0,width=\textwidth]{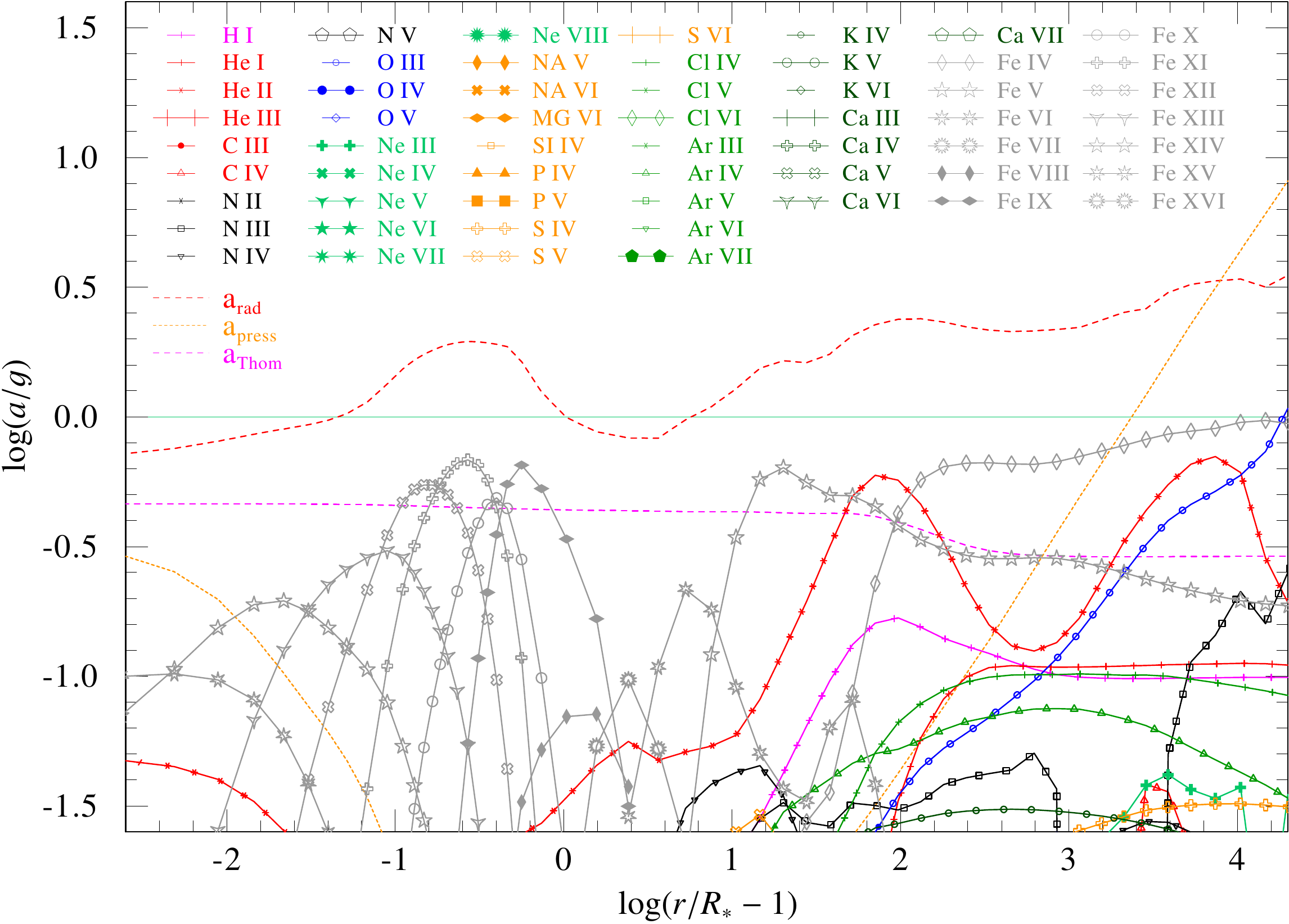}
     \caption{Contributions to the radiative acceleration in a WN star with $T_\mathrm{eff}(\tau = 2/3) = 32\,$kK, $T_\mathrm{eff}(\tau_\mathrm{sonic} = 27) = 122\,$kK, and $X_\mathrm{H} = 0.2$.
		          The underlying model to produce this plot is taken from Sander et al. (2022, submitted).}
   \label{fig:leadions-wnh}
\end{center}
\end{figure}

The contributions of the individual elements to the total radiative acceleration is not trivial. Early calculations for $\zeta$\,Pup and $\tau$\,Sco by \citet{Pauldrach1987} demonstrated that the contributions from different elements to the total (modified) CAK force multiplier $\mathcal{M}$ are substantially changing between different types of stars and different parts of the wind. Moreover, elements not visible in the observable spectra like Ne and Ar can be efficient wind drivers despite their low abundances.

With the detailed CMF radiative transfer we can nowadays obtain an even more detailed picture, which we illustrate in Figs.\,\ref{fig:leadions-osg}, \ref{fig:leadions-bsg} and \ref{fig:leadions-wnh} for three different types of hot stars. In these plots, we only separate Thomson scattering and gas pressure, while the individual ion contributions include both line and (true) continuum contributions. The comparison of the three plots underlines the importance of Fe for driving stellar winds. As briefly mentioned above, iron is particularly crucial for launching a stellar wind in the first place and thereby indirectly determining the mass-loss rate $\dot{M}$ \citep[e.g.][]{Pauldrach+1993,deKoter+1997,Vink+1999}. In the outer wind, a variety of elements is responsible for the further wind acceleration, thereby setting $v_\infty$. Especially the latter includes elements that are usually not detectable in the spectra, such as argon in the case of the depicted O supergiant model (Fig.\,\ref{fig:leadions-osg}).  
In general, the elemental abundance is not a good indicator for its contribution to the wind driving. In the B supergiant model (Fig.\,\ref{fig:leadions-bsg}), ions of Fe, Si, and S  dominate the acceleration of the outer wind, despite the more abundant CNO elements. Hydrogen and helium do not contribute notably to the line acceleration. However, their bound-free contribution can be significant if their ionization stage changes in the wind. This is the case in the depicted WR model (Fig.\,\ref{fig:leadions-wnh}), where the contributions of He\,\textsc{ii} and H\,\textsc{i} in the outer wind are clearly visible.

The more dense winds of WR stars are still quite enigmatic, but the example in Fig.\,\ref{fig:leadions-wnh} illustrates that for He-burning stars the wind launching region is much further inward than in the case of O and B stars. The launching of the wind is then fully tied to the iron opacities \citep[see also][]{GraefenerHamann2005,VinkdeKoter2005}. The observational evidence for WR stars appearing preferably in higher metallicity environments \citep[e.g.][]{Shenar+2020} is thus nicely backed by wind driving studies. This means that also the mass-loss of WR stars scales (mainly) with iron, which has severe consequences for stellar evolution in the early Universe \citep[e.g.][]{Vink+2021}. In older population synthesis \citep[e.g.][]{Hurley+2000}, WR winds were treated as metallicity-independent due to the WC stage where self-produced carbon reaches the surface of the stars. However, as indicated by the modelling results in \citet{Sander+2020}, the higher ionization stages of carbon do not yield significant line opacities and thus cannot compensate the declining impact of iron opacities on the wind mass-loss rates at lower metallicities.

Beside the classical, He-burning WR stars, the WR phenomenon also occurs at the upper end of the main sequence. Due to their high $L/M$-ratio, very massive stars (VMS) with $M_\mathrm{ini} \approx 100\,M_\odot$ and higher already show WN-type spectra during their central hydrogen burning stage. Such objects, which exist for example in NGC\,3603 and the R136 cluster in the LMC \citep[e.g.][]{CrowtherDessart1998} or the Galactic Center region \citep[e.g.][]{Figer+2002}, have typically less dense winds than the bulk of the He-burning WR stars (if compared at the same metallicity). In the last 15 years, these stars have gained substantial attention, not least due to their influence on stellar populations \citep[e.g.][]{Vink+2015,Senchyna+2021}. Yet, major differences exist in modelling their mass-loss rates \citep[e.g.][]{GraefenerHamann2008,Vink+2011,Graefener2021} and the resulting evolutionary paths \citep[e.g.,][see also these proceedings]{Sabhahit+2022}.

\section{The influence of companions}
  \label{sec:companions}
	
Multiplicity is common among massive stars. Consequently, a significant fraction of massive stars has one or more stellar or compact companion. Depending on the configuration of the system -- concerning both the properties of the objects as well as their orbital parameters -- the (radiative) driving of the stellar winds can be considerably affected. In any case, the presence of a companion breaks the spherical symmetry of the system. While this is also the case for rotating individual stars \citep[e.g.][]{MuellerVink2014}, the geometry is not as complex as in case of a companion. With additional radiation coming from non-radial directions or radiation missing from particular angles, typical assumptions about the radiation field inherent to most 1D model efforts are invalid. 2D and 3D calculations become necessary, but are numerically much more costly unless other modelling aspects are significantly simplified. For practical reasons, a full 3D radiative transfer is typically limited to specific applications \cite[e.g.][]{Surlan+2013,Hennicker+2018,Hennicker+2020} without performing a full non-LTE model atmosphere calculations, expect for first test cases \citep[][]{HauschildtBaron2014}. Multi-dimensional radiative driving studies incorporating the influence of companions therefore need to make significant approximations either the radiative acceleration or for the influence of the companion. Without claiming completeness, the following aspects can lead to major changes in radiation-driven winds of hot stars:

\begin{itemize}
  \item[\emph{Binary evolution:}] Mass transfer in a binary system considerably changes the stellar parameters of both components. In particular in cases where the donor star is already evolved, the loss of surface material leads to a configuration that fosters stronger stellar winds. Directly, the higher $L/M$-ratio of the donor star, which is now ``overluminous'' compared to a single star of the same mass, brings the donor closer to the Eddington limit and thus increases the mass-loss rate. In addition, the mass transfer can prevent the donor from leaving the regime of radiation-driven winds due to avoiding cooler surface temperatures.
  \item[\emph{Roche Lobe modification:}] The existence of a stellar wind with $\Gamma_\mathrm{rad} > 0$ modifies the effective gravity $g_\mathrm{eff} = g (1 - \Gamma_\mathrm{rad})$ of a star. Hot star winds eventually reach supersonic speeds and thus $\Gamma_\mathrm{rad} > 1$, meaning that the effective gravity becomes negative and there is no longer a Roche Lobe for this star \citep[e.g.][]{Dermine+2009}. Unless the subsonic layers of the star alone fill their Roche Lobe, also the term ``Roche Lobe overflow'' (RLOF) becomes meaningless. Due to the absence of a Roche Lobe, material can escape in any direction. Still, the gravitational potential of the companion provides a preference for matter overflow, leading to a so-called ``focussed wind'' scenario \citep{FriendCastor1982,GiesBolton1986,HiraiMandel2021}. Recently, the term ``wind RLOF'' has also been used in the context of hot star winds to describe supersonic winds focussed towards an accretor \citep[][]{ElMellah+2019a,ElMellah+2019b}. This terminology was originally introduced by \citet{MohamedPodsiadlowski2007} for AGB and RSG stars with strong winds. There, the stars as such do not fill their Roche Lobe. However, as RSG and AGB winds are characterized by subsonic velocities, the wind material can fill the star's Roche Lobe in a similar way than the (hydrostatic) star itself. The material is then removed mainly through the inner Lagrangian point. Unless the acceleration is extremely shallow, a similar situation of nearly complete redirection onto the accretor does not occur for hot star winds. \citet{ElMellah+2019b} calculated that a maximum of approximately 20\% of the donor wind could be accreted by the (compact) companion in an X-ray binary. 
  \item[\emph{Wind-wind collision:}] In typical cases, both companions in a massive binary system are hot stars, leading e.g.\ to systems of type O+O or WR+O. The collision region of the winds from the two components then gives rise to a variety of phenomena, most notably (additional) X-ray emission arising from the shock region and non-thermal radio emission \citep[e.g.][]{PrilutskiiUsov1976,Stevens+1992,DeBecker2007}. The additional X-ray flux can ionize the wind and thus alter the radiative acceleration. In most cases, the higher ionization stages will provide fewer line transitions, thus leading to a reduction of the wind driving. A further reduction of the wind driving arises due to the existence of the UV-intensive companion itself. The absorption of the photons from the other hot star can lead to the so-called \textit{radiative breaking} \citep[e.g.][]{StevensPollock1994,Gayley+1997} as the absorbed momentum has a component opposite to the radial wind direction. In extreme cases, this effect could lead to a collapse of the wind-colliding region, enabling temporary accretion of material from one star onto the other \citep[e.g.][]{Soker2007}.
More frequently, however, are other wind collision effects, such as (optical) excess line emission in WR+O binaries \citep[e.g.][]{Luehrs1997,Hill+2002} or dust production, in particular in environments where the wind from a carbon-rich WR star (WC star) meets a hydrogen-rich O-star wind \citep[e.g.][]{Allen+1972,Williams+1990,Lau+2021}.  		
  \item[\emph{Irradiation:}] Massive binaries including a compact object (i.e. a neutron star or black hole) as one of their components, can be a major X-ray source due to the accretion of material onto the compact object. In these so-called High Mass X-ray binaries (HMXBs), the ionization structure and thus also the driving of the stellar wind of the non-degenerate star can be significantly altered if the X-ray source is close enough. Orbital modulations of characteristic UV wind profiles were first detected in the late 1970s by \citet{HatchettMcCray1977}, providing observational evidence for the phase-dependent change of the donor star winds. Similar to the X-ray effect in colliding winds, the higher ionization leads to a reduction or termination of the radiative driving \citep[e.g.][]{MacGregorVitello1982,StevensKallman1990}. More detailed calculations of the radiative driving reveal that contrary to the general reduction, there can also be an enhancement of the radiative force in some cases if the X-ray luminosity remains moderate \citep[e.g.][]{Krticka+2012,Sander+2018}. The lower terminal wind velocities resulting from X-ray irradiation can foster the accretion of material onto the compact object, thereby affecting the further evolution of the system \citep[e.g.][]{Krticka+2022,Ramachandran+2022}. 
\end{itemize}	

\section{Summary \& Conclusions}
  \label{sec:summary}

  The outermost layers of hot stars provide an inherent non-LTE environment. The winds of these stars are mainly driven by radiative acceleration. The most important contributors to this acceleration are Thomson scattering and the absorption of photons in spectral lines. As the latter corresponds to a strongly radius-dependent opacity and the launch of the winds is usually impossible without considering this additional contribution, hot star winds are often simply termed \emph{line-driven}. In reality, additional acceleration components can contribute, such as free-free and bound-free opacities, but their importance is usually much weaker than the line and free electron opacities. Gas pressure only has a substantial impact in the subsonic layers. 
	
	The line driving opacity itself consists of a plethora of contributing elements and ions with individual contributions depending strongly on the particular stellar parameters, most notably $T_\mathrm{eff}$. The iron group elements play a crucial role in hot star winds due to their millions of available line transitions and the resulting supply of opacity (and thus acceleration). In the onset region of hot star winds, lighter elements are often too highly ionized to contribute significant line opacities. Thus, it is the iron opacity that largely determines the mass-loss rate of hot star winds. Overall, one can distinguish between two types of line-driven hot star winds:
	
\begin{itemize}
  \item[\emph{OB-type winds}] are optically thin with individual optically thick lines. The imprint of the stellar wind is mainly seen in the UV with additional, smaller features in the optical and (near) IR for higher mass-loss rates (e.g. in supergiants). OB-type winds occur in most hot, massive stars, ranging from the main sequence over the supergiant regime to hydrogen-stripped stars that are not sufficiently close to the Eddington limit to launch WR-type winds.
  \item[\emph{WR-type winds}] are optically thick up to large radii. Their optical spectra are characterized by emission lines. Caused by the proximity to the Eddington limit, WR-type winds occur not only in evolved hydrogen-depleted stars (classical WR stars), but also in very massive (hydrogen-rich) stars and (some) luminous blue variables such as AG\,Car, $\eta$\,Car, or P\,Cyg. 
\end{itemize}
	
Radiatively driven winds are commonly described and simulated with either the semi-analytic (modified) CAK description or more comprehensive Monte Carlo or CMF calculations. The latter two allow for a more realistic physical treatment. In particular, the CMF modelling can provide a local consistency, resulting in complex shapes for the flux-weighted mean opacity $\varkappa_F$ that cannot be sufficiently approximated by using Rosseland opacities. While recipes for the wind mass-loss rate $\dot{M}$ based on such detailed $\varkappa_F$-models become more and more available, the aim for an analytic description usually requires compromises. Consequently, no current $\dot{M}$-formula captures the full $\varkappa_F$-complexity.

Companions of hot stars can have a severe influence on their radiatively driven stellar winds. Via mass-transfer in binary systems, they can cause stronger stellar winds for one of the stars. X-rays due to wind-wind collision or accretion can significantly alter the ionization stratification, usually weakening the radiative acceleration. The supersonic nature of hot star winds further modifies the effective gravity of the stars, making it difficult or even impractical to define a Roche lobe.

Radiatively-driven winds are subject to inherent instabilities, requiring time-dependent modelling to study their effects. While considerable imprints of time-dependent processes, e.g. clumping and shocks, are seen in such simulations, they also show that the bulk of the matter outflow can be sufficiently described by stationary models. Quantitative spectroscopy of hot stars makes use of this result in so-called unified model atmospheres where the hydrostatic and the supersonic layers are treated consistently to obtain synthetic spectra of hot stars including their winds. Observational evidence indicates that both OB- and WR-type winds are inhomogeneous (``clumped''). The origin of the clumps is debated and might even reach down into the subsonic layers. The existence of clumps requires approximations in (1D) model atmospheres and considerably impacts the diagnosis of wind mass-loss rates.

\section*{Acknowledgements}

AACS is funded by the Deutsche Forschungsgemeinschaft (DFG, German Research Foundation) in the form of an Emmy Noether Research Group -- Project-ID 445674056 (SA4064/1-1, PI Sander)

\def\apj{{ApJ}}    
\def\nat{{Nature}}    
\def\jgr{{JGR}}    
\def\apjl{{ApJ Letters}}    
\def\aap{{A\&A}}   
\def\aaps{{A\&A Supplement}}  
\def\aapr{{The Astronomy and Astrophysics Review}} 
\def\mnras{{MNRAS}}
\def\aj{{AJ}}
\def\pasa{{PASA}}
\def\pasp{{PASP}}
\def\ssr{{Space Science Reviews}}
\def\sovast{{Soviet Astronomy}}
\let\mnrasl=\mnras



\begin{thebibliography}{}

\bibitem[Abbott(1982)]{Abbott1982}
{Abbott, D.~C.} 1982,
\textit{\apj}, 259, 282

\bibitem[Abbott \& Lucy(1985)]{AbbottLucy1985}
{Abbott, D.~C. \& Lucy, L.~B.} 1985,
\textit{\apj}, 288, 679

\bibitem[Allen et al.(1972)]{Allen+1972} 
{Allen, D.~A., Swings, J.~P., \& Harvey, P.~M.} 1972, 
\textit{\aap}, 20, 333

\bibitem[Baron \& Hauschildt(1998)]{BaronHauschildt1998} 
{Baron, E. \& Hauschildt, P.~H.} 1998, 
\textit{\apj}, 495, 370

\bibitem[Beals(1929)]{Beals1929} 
{Beals, C.~S.} 1929, 
\textit{\mnras}, 90, 202

\bibitem[Bj{\"o}rklund et al.(2021)]{Bjoerklund+2021} 
{Bj{\"o}rklund, R., Sundqvist, J.~O., Puls, J., et al.} 2021, 
\textit{\aap}, 648, A36

\bibitem[Carlberg(1980)]{Carlberg1980} 
{Carlberg, R.~G.} 1980, 
\textit{\apj}, 241, 1131

\bibitem[Castor(1974)]{Castor1974}
{Castor, J.~L.} 1974,
\textit{\mnras}, 169, 279

\bibitem[Castor et al.(1975)]{Castor+1975}
{Castor, J.~I., Abbott, D.~C., \& Klein, R.~I.} 1975,
\textit{\apj}, 195, 157

\bibitem[Cantiello et al.(2009)]{Cantiello+2009} 
{Cantiello, M., Langer, N., Brott, I., et al.} 2009, 
\textit{\aap}, 499, 279

\bibitem[Crowther \& Dessart(1998)]{CrowtherDessart1998} 
{Crowther, P.~A. \& Dessart, L.} 1998, 
\textit{\mnras}, 296, 622

\bibitem[De Becker(2007)]{DeBecker2007} 
{De Becker, M.} 2007, 
\textit{\aapr}, 14, 171

\bibitem[de Koter et al.(1997)]{deKoter+1997}
{de Koter, A., Heap, S.~R., \& Hubeny, I.} 1997, 
\textit{\apj}, 477, 792

\bibitem[Dermine et al.(2009)]{Dermine+2009} 
{Dermine, T., Jorissen, A., Siess, L., et al.} 2009, 
\textit{\aap}, 507, 891

\bibitem[Dessart \& Owocki(2005)]{DessartOwocki2005} 
{Dessart, L. \& Owocki, S.~P.} 2005, 
\textit{\aap}, 437, 657

\bibitem[Driessen et al.(2021)]{Driessen+2021}
{Driessen, F.~A., Kee, N.~D., \& Sundqvist, J.~O.} 2021, 
\textit{\aap}, 656, A131

\bibitem[Driessen et al.(2022)]{Driessen+2022} 
{Driessen, F.~A., Sundqvist, J.~O., \& Dagore, A.} 2022, 
\textit{\aap}, 663, A40

\bibitem[El Mellah et al.(2019a)]{ElMellah+2019a} 
{El Mellah, I., Sander, A.~A.~C., Sundqvist, J.~O., et al.} 2019, 
\textit{\aap}, 622, A189

\bibitem[El Mellah et al.(2019b)]{ElMellah+2019b} 
{El Mellah, I., Sundqvist, J.~O., \& Keppens, R.} 2019,
\textit{\aap}, 622, L3

\bibitem[Feldmeier et al.(1997)]{Feldmeier+1997} 
{Feldmeier, A., Puls, J., \& Pauldrach, A.~W.~A.} 1997,
\textit{\aap}, 322, 878

\bibitem[Figer et al.(2002)]{Figer+2002} 
{Figer, D.~F., Najarro, F., Gilmore, D., et al.} 2002, 
\textit{\apj}, 581, 258

\bibitem[Friend \& Castor(1982)]{FriendCastor1982}
{Friend, D.~B. \& Castor, J.~I.} 1982, 
\textit{\apj}, 261, 293

\bibitem[Friend \& Abbott(1986)]{FriendAbbott1986}
{Friend, D.~B. \& Abbott, D.~C.} 1986,
\textit{\apj}, 311, 701

\bibitem[Gayley(1995)]{Gayley1995} 
{Gayley, K.~G.} 1995, 
\textit{\apj}, 454, 410

\bibitem[Gayley et al.(1997)]{Gayley+1997} 
{Gayley, K.~G., Owocki, S.~P., \& Cranmer, S.~R.} 1997, 
\textit{\apj}, 475, 786

\bibitem[Gormaz-Matamala et al.(2021)]{Gormaz-Matamala+2021} 
{Gormaz-Matamala, A.~C., Cur{\'e}, M., Hillier, D.~J., et al.} 2021, 
\textit{\apj}, 920, 64

\bibitem[Gies \& Bolton(1986)]{GiesBolton1986} 
{Gies, D.~R. \& Bolton, C.~T.} 1986,
\textit{\apj}, 304, 389

\bibitem[Gr{\"a}fener \& Hamann(2005)]{GraefenerHamann2005} 
{Gr{\"a}fener, G. \& Hamann, W.-R.} 2005, 
\textit{\aap}, 432, 633

\bibitem[Gr{\"a}fener \& Hamann(2008)]{GraefenerHamann2008} 
{Gr{\"a}fener, G. \& Hamann, W.-R.} 2008, 
\textit{\aap}, 482, 945

\bibitem[Gr{\"a}fener(2021)]{Graefener2021} 
{Gr{\"a}fener, G.} 2021, 
\textit{\aap}, 647, A13

\bibitem[Hamann et al.(1991)]{Hamann+1991} 
{Hamann, W.-R., Duennebeil, G., Koesterke, L., et al.} 1991, 
\textit{\aap}, 249, 443

\bibitem[Hatchett \& McCray(1977)]{HatchettMcCray1977} 
{Hatchett, S. \& McCray, R.} 1977, 
\textit{\apj}, 211, 552

\bibitem[Hauschildt \& Baron(2014)]{HauschildtBaron2014} 
{Hauschildt, P.~H. \& Baron, E.} 2014, 
\textit{\aap}, 566, A89

\bibitem[Hawcroft et al.(2021)]{Hawcroft+2021} 
{Hawcroft, C., Sana, H., Mahy, L., et al.} 2021, 
\textit{\aap}, 655, A67

\bibitem[Hennicker et al.(2018)]{Hennicker+2018} 
{Hennicker, L., Puls, J., Kee, N.~D., et al.} 2018, 
\textit{\aap}, 616, A140

\bibitem[Hennicker et al.(2020)]{Hennicker+2020} 
{Hennicker, L., Puls, J., Kee, N.~D., et al.} 2020, 
\textit{\aap}, 633, A16

\bibitem[Holzer(1977)]{Holzer1977} 
{Holzer, T.~E.} 1977,
\textit{\jgr}, 82, 23

\bibitem[Hill et al.(2002)]{Hill+2002} 
{Hill, G.~M., Moffat, A.~F.~J., \& St-Louis, N.} 2002, 
\textit{\mnras}, 335, 1069

\bibitem[Hillier \& Miller(1998)]{HillierMiller1998} 
{Hillier, D.~J. \& Miller, D.~L.} 1998,
\textit{\apj}, 496, 407

\bibitem[Hirai \& Mandel(2021)]{HiraiMandel2021} 
{Hirai, R. \& Mandel, I.} 2021, 
\textit{\pasa}, 38, e056

\bibitem[Hurley et al.(2000)]{Hurley+2000} 
{Hurley, J.~R., Pols, O.~R., \& Tout, C.~A.} 2000, 
\textit{\mnras}, 315, 543

\bibitem[Iglesias \& Rogers(1996)]{IglesiasRogers1996}
{Iglesias, C.~A. \& Rogers, F.~J.} 1996,
\textit{\apj}, 464, 943

\bibitem[Krti{\v{c}}ka \& Kub{\'a}t(2010)]{KrtickaKubat2010} 
{Krti{\v{c}}ka, J. \& Kub{\'a}t, J.} 2010, 
\textit{\aap}, 519, A50

\bibitem[Krti{\v{c}}ka et al.(2012)]{Krticka+2012} 
{Krti{\v{c}}ka, J., Kub{\'a}t, J., \& Skalick{\'y}, J.} 2012, 
\textit{\apj}, 757, 162

\bibitem[Krti{\v{c}}ka \& Kub{\'a}t(2017)]{KrtickaKubat2017} 
{Krti{\v{c}}ka, J. \& Kub{\'a}t, J.} 2017, 
\textit{\aap}, 606, A31

\bibitem[Krti{\v{c}}ka \& Kub{\'a}t(2018)]{KrtickaKubat2018} 
{Krti{\v{c}}ka, J. \& Kub{\'a}t, J.} 2018, 
\textit{\aap}, 612, A20

\bibitem[Krti{\v{c}}ka et al.(2022)]{Krticka+2022} 
{Krti{\v{c}}ka, J., Kub{\'a}t, J., \& Krti{\v{c}}kov{\'a}, I.} 2022, 
\textit{\aap}, 659, A117

\bibitem[Kudritzki et al.(1989)]{Kudritzki+1989} 
{Kudritzki, R.~P., Pauldrach, A., Puls, J., et al.} 1989,
\textit{\aap}, 219, 205

\bibitem[Lamers et al.(1995)]{Lamers+1995}
{Lamers, H.~J.~G.~L.~M., Snow, T.~P., \& Lindholm, D.~M.} 1995, 
\textit{\apj}, 455, 269

\bibitem[L{\'e}pine \& Moffat(2008)]{LepineMoffat2008} 
{L{\'e}pine, S. \& Moffat, A.~F.~J.} 2008, 
\textit{\aj}, 136, 548

\bibitem[Lau et al.(2021)]{Lau+2021} 
{Lau, R.~M., Hankins, M.~J., Kasliwal, M.~M., et al.} 2021, 
\textit{\apj}, 909, 113

\bibitem[Lucy \& Solomon(1967)]{LucySolomon1967}
{Lucy, L.~B. \& Solomon, P.~M.} 1967,
\textit{\aj}, 72, 310

\bibitem[Lucy \& Solomon(1970)]{LucySolomon1970} 
{Lucy, L.~B. \& Solomon, P.~M.} 1970,
\textit{\apj}, 159, 879

\bibitem[Lucy(2010)]{Lucy2010} 
{Lucy, L.~B.} 2010, 
\textit{\aap}, 524, A41

\bibitem[L{\"u}hrs(1997)]{Luehrs1997} 
{L{\"u}hrs, S.} 1997, 
\textit{\pasp}, 109, 504

\bibitem[MacGregor et al.(1979)]{MacGregor+1979} 
{MacGregor, K.~B., Hartmann, L., \& Raymond, J.~C.} 1979, 
\textit{\apj}, 231, 514

\bibitem[MacGregor \& Vitello(1982)]{MacGregorVitello1982} 
{MacGregor, K.~B. \& Vitello, P.~A.~J.} 1982, 
\textit{\apj}, 259, 267

\bibitem[Milne(1926)]{Milne1926} 
{Milne, E.~A.} 1926,
\textit{\mnras}, 86, 459

\bibitem[Mihalas et al.(1975)]{Mihalas+1975} 
{Mihalas, D., Kunasz, P.~B., \& Hummer, D.~G.} 1975, 
\textit{\apj}, 202, 465

\bibitem[Moens et al.(2022)]{Moens+2022} 
{Moens, N., Poniatowski, L.~G., Hennicker, L., et al.} 2022, 
\textit{\aap}, 665, A42

\bibitem[Mohamed \& Podsiadlowski(2007)]{MohamedPodsiadlowski2007} 
{Mohamed, S. \& Podsiadlowski, P.} 2007, 
\textit{15th European Workshop on White Dwarfs}, 372, 397

\bibitem[Morton(1967)]{Morton1967} 
{Morton, D.~C.} 1967, 
\textit{\apj}, 150, 535

\bibitem[M{\"u}ller \& Vink(2008)]{MuellerVink2008} 
{M{\"u}ller, P.~E. \& Vink, J.~S.} 2008, 
\textit{\aap}, 492, 493

\bibitem[M{\"u}ller \& Vink(2014)]{MuellerVink2014} 
{M{\"u}ller, P.~E. \& Vink, J.~S.} 2014, 
\textit{\aap}, 564, A57

\bibitem[Owocki \& Rybicki(1984)]{OwockiRybicki1984} 
{Owocki, S.~P. \& Rybicki, G.~B.} 1984,
\textit{\apj}, 284, 337

\bibitem[Owocki \& Puls(1999)]{OwockiPuls1999} 
{Owocki, S.~P. \& Puls, J.} 1999, 
\textit{\apj}, 510, 355

\bibitem[Pauldrach et al.(1986)]{Pauldrach+1986} 
{Pauldrach, A., Puls, J., \& Kudritzki, R.~P.} 1986,
\textit{\aap}, 164, 86

\bibitem[Pauldrach(1987)]{Pauldrach1987} 
{Pauldrach, A.} 1987,
\textit{\aap}, 183, 295

\bibitem[Pauldrach \& Puls(1990)]{PauldrachPuls1990} 
{Pauldrach, A.~W.~A. \& Puls, J.} 1990, 
\textit{\aap}, 237, 409

\bibitem[Pauldrach et al.(1993)]{Pauldrach+1993} 
{Pauldrach, A.~W.~A., Feldmeier, A., Puls, J., et al.} 1993, 
\textit{\ssr}, 66, 105

\bibitem[Prilutskii \& Usov(1976)]{PrilutskiiUsov1976} 
{Prilutskii, O.~F. \& Usov, V.~V.} 1976, 
\textit{\sovast}, 20, 2

\bibitem[Poniatowski et al.(2022)]{Poniatowski+2022} 
{Poniatowski, L.~G., Kee, N.~D., Sundqvist, J.~O., et al.} 2022, 
\textit{\aap}, 667, A113, arXiv:2204.09981

\bibitem[Puls et al.(1993)]{Puls+1993} 
{Puls, J., Pauldrach, A.~W.~A., Kudritzki, R.-P., et al.} 1993, 
\textit{Reviews in Modern Astronomy}, 6, 271

\bibitem[Puls et al.(2000)]{Puls+2000} 
{Puls, J., Springmann, U., \& Lennon, M.} 2000, 
\textit{\aaps}, 141, 23

\bibitem[Puls et al.(2020)]{Puls+2020} 
{Puls, J., Najarro, F., Sundqvist, J.~O., et al.} 2020, 
\textit{\aap}, 642, A172

\bibitem[Ramachandran et al.(2019)]{Ramachandran+2019} 
{Ramachandran, V., Hamann, W.-R., Oskinova, L.~M., et al.} 2019, 
\textit{\aap}, 625, A104

\bibitem[Ramachandran et al.(2022)]{Ramachandran+2022} 
{Ramachandran, V., Oskinova, L.~M., Hamann, W.-R., et al.} 2022, 
\textit{\aap}, 667, A77, arXiv:2208.07773

\bibitem[Sabhahit et al.(2022)]{Sabhahit+2022} 
{Sabhahit, G.~N., Vink, J.~S., Higgins, E.~R., et al.} 2022, 
\textit{\mnras}, 514, 3736

\bibitem[Sander et al.\,(2017)]{Sander+2017} 
{Sander, A.~A.~C., Hamann, W.-R., Todt, H., et al.} 2017, 
\textit{\aap}, 603, A86

\bibitem[Sander et al.\,(2018)]{Sander+2018} 
{Sander, A.~A.~C., F{\"u}rst, F., Kretschmar, P., et al.} 2018,
\textit{\aap}, 610, A60

\bibitem[Sander et al.(2020)]{Sander+2020} 
{Sander, A.~A.~C., Vink, J.~S., \& Hamann, W.-R.} 2020, 
\textit{\mnras}, 491, 4406

\bibitem[Sander \& Vink(2020)]{SanderVink2020} 
{Sander, A.~A.~C. \& Vink, J.~S.\ 2020}, 
\textit{\mnras}, 499, 873

\bibitem[Schmutz (1993)]{Schmutz1993}
{Schmutz, W.} 1993,
\textit{\ssr}, 66, 253

\bibitem[Senchyna et al.(2021)]{Senchyna+2021} 
{Senchyna, P., Stark, D.~P., Charlot, S., et al.} 2021, 
\textit{\mnras}, 503, 6112

\bibitem[Shenar et al.(2020)]{Shenar+2020} 
{Shenar, T., Gilkis, A., Vink, J.~S., et al.} 2020,
\textit{\aap}, 634, A79

\bibitem[Sobolev(1960)]{Sobolev1960} 
{Sobolev, V.~V.} 1960, 
\textit{Harvard University Press}

\bibitem[Soker(2007)]{Soker2007} 
{Soker, N.} 2007, 
\textit{\apj}, 661, 482

\bibitem[Stevens \& Kallman(1990)]{StevensKallman1990} 
{Stevens, I.~R. \& Kallman, T.~R.} 1990, 
\textit{\apj}, 365, 321

\bibitem[Stevens et al.(1992)]{Stevens+1992} 
{Stevens, I.~R., Blondin, J.~M., \& Pollock, A.~M.~T.} 1992, 
\textit{\apj}, 386, 265

\bibitem[Stevens \& Pollock(1994)]{StevensPollock1994} 
{Stevens, I.~R. \& Pollock, A.~M.~T.} 1994, 
\textit{\mnras}, 269, 226

\bibitem[Sundqvist \& Owocki(2013)]{SundqvistOwocki2013} 
{Sundqvist, J.~O. \& Owocki, S.~P.} 2013, 
\textit{\mnras}, 428, 1837

\bibitem[Sundqvist et al.(2019)]{Sundqvist+2019} 
{Sundqvist, J.~O., Bj{\"o}rklund, R., Puls, J., et al.} 2019,
\textit{\aap}, 632, A126

\bibitem[{\v{S}}urlan et al.(2012)]{Surlan+2012} 
{\v{S}urlan, B., Hamann, W.-R., Kub{\'a}t, J., et al.} 2012, 
\textit{\aap}, 541, A37

\bibitem[{\v{S}}urlan et al.(2013)]{Surlan+2013} 
{\v{S}urlan, B., Hamann, W.-R., Aret, A., et al.} 2013, 
\textit{\aap}, 559, A130

\bibitem[ud-Doula \& Owocki(2002)]{udDoulaOwocki2002} 
{ud-Doula, A. \& Owocki, S.~P.} 2002, 
\textit{\apj}, 576, 413

\bibitem[Vink et al.(1999)]{Vink+1999} 
{Vink, J.~S., de Koter, A., \& Lamers, H.~J.~G.~L.~M.} 1999, 
\textit{\aap}, 350, 181

\bibitem[Vink et al.(2000)]{Vink+2000} 
{Vink, J.~S., de Koter, A., \& Lamers, H.~J.~G.~L.~M.} 2000, 
\textit{\aap}, 362, 295

\bibitem[Vink et al.(2001)]{Vink+2001} 
{Vink, J.~S., de Koter, A., \& Lamers, H.~J.~G.~L.~M.} 2001, 
\textit{\aap}, 369, 574

\bibitem[Vink \& de Koter(2005)]{VinkdeKoter2005} 
{Vink, J.~S. \& de Koter, A.} 2005, 
\textit{\aap}, 442, 587

\bibitem[Vink et al.(2011)]{Vink+2011} 
{Vink, J.~S., Muijres, L.~E., Anthonisse, B., et al.} 2011, 
\textit{\aap}, 531, A132

\bibitem[Vink et al.(2015)]{Vink+2015} 
{Vink, J.~S., Heger, A., Krumholz, M.~R., et al.} 2015, 
\textit{Highlights of Astronomy}, 16, 51

\bibitem[Vink et al.(2021)]{Vink+2021} 
{Vink, J.~S., Higgins, E.~R., Sander, A.~A.~C., et al.} 2021, 
\textit{\mnras}, 504, 146

\bibitem[Vink \& Sander(2021)]{VinkSander2021} 
{Vink, J.~S. \& Sander, A.~A.~C.} 2021, 
\textit{\mnras}, 504, 2051
	
\bibitem[Williams et al.(1990)]{Williams+1990} 
{Williams, P.~M., van der Hucht, K.~A., The, P.~S., et al.} 1990, 
\textit{\mnras}, 247, 18P	
	
\end{thebibliography}
\end{document}